%% file: main.tex
\documentclass{optica-article}

\journal{opticajournal} 

\articletype{Research Article}

\usepackage{lineno}

\usepackage{siunitx}

\begin{document}


\title{A Finite-Difference Time-Domain approach for dispersive magnetic media}

\author{Jasmin Graf, \authormark{1,2,3,4*} Joshua Baxter, \authormark{4,5}  Sanchar Sharma, \authormark{1,2} Silvia {Viola Kusminskiy}, \authormark{1,2,4} Lora Ramunno, \authormark{4,5,6}}

\address{\authormark{1}Institut f{\"u}r Theoretische Festk{\"o}rperphysik, RWTH Aachen University, 52056 Aachen, Germany\\
\authormark{2}Max Planck Institute for the Science of Light, 91058 Erlangen, Germany\\
\authormark{3}Friedrich-Alexander-Universit{\"a}t Erlangen-N{\"u}rnberg, Department of Physics, 91058 Erlangen, Germany\\
\authormark{4}Max Planck-uOttawa Centre for Extreme and Quantum Photonics, University of Ottawa, ON, K1N 6N5 Ottawa, Canada\\
\authormark{5}Department of Physics, University of Ottawa, ON, K1N 6N5 Ottawa, Canada\\
\authormark{6}Nexus for Quantum Technologies Institute, University of Ottawa, ON, K1N 6N5 Ottawa, Canada\\

}

\email{\authormark{*}jasmin.graf@mpl.mpg.de}


\begin{abstract*}
We extend the Finite-Difference Time-Domain method to treat dispersive magnetic media by incorporating magneto-optical effects through a frequency-dependent permittivity tensor. For benchmarking our method, we consider the light scattering on a magnetic sphere in the Mie regime.
We first derive the analytical scattering expressions which predict a peak broadening in the scattering efficiency due to the atomic energy level splitting in the presence of a magnetic field, together with an additional rotated part in the scattered field profile due to the Faraday rotation.
We show that our numerical method is able to capture the main scattering features and discuss its limitations and possible improvements in accuracy. 
\end{abstract*}


\section{Introduction}
Magneto-optical (MO) materials, in which the magnetization affects the polarization of optical photons, are vital for non-reciprocal optical technologies such as circulators and isolators~\cite{MO_Roadmap}. Garnets (and related heterostructures) are especially promising for MO devices, owing to their strong magneto-optical activity combined with low losses \cite{rizal_MO_Rev,Li_MO_YIG,onbasli_CeYIG,zhu_IFE,fakhrul_BiYIG,nur-e-alam_BiYIG}. Strong MO effects also enable information exchange between the photons and the magnetization, via Brillouin light scattering (BLS). BLS is an extremely sensitive probe of magnetization~\cite{BLS_Rev1,BLS_Rev2} with a high spatial and temporal resolution. Such an information exchange is promising for the coveted microwave-to-optical conversion~\cite{MtoO_magnons, hybrid_systems_6c, engelhardt_2022, optomagnonic_new_4} since the magnetization couples strongly to microwaves \cite{tabuchi_coherent_2015,mw-magnonics_zhang,mw-magnonics_haigh}. This conversion would require a strong and coherent coupling between the magnetization and optical photons.

This coupling can be boosted by an optical cavity, where the dielectric material is patterned to serve as a resonator~\cite{optomagnonics_review, optomagnonics_kusminskiy, optomagnonics_hisatomi, optomagnonics_bittencourt, optomagnonics_sharma_2}. Examples include magnetic spheres~\cite{optomagnonics_sharma, optomagnonics_osada, optomagnonics_osada_2, optomagnonics_osada_3, optomagnonics_haigh, optomagnonics_haigh_2, optomagnonics_haigh_4, optomagnonics_haigh_5, optomagnonics_zhang, lachance-quirion_resolving_2016, optomagnonics_wachter, magnetic_sphere_4, magnetic_sphere_5}, slabs~\cite{optomagnonics_liu}, disks~\cite{OM_disk}, waveguides~\cite{optomagnonics_zhu, optomagnonic_new_4, optomagnonics_yu}, layered structures~\cite{optomagnonic_new_2, optomagnonic_new_1, optomagnonics_haigh_5, optomagnonics_haigh_3, layered_Pantazopoulos_2019, layered_Pantazopoulos:18}, and crystals~\cite{OM-crystal}.
Further developments investigate optical cavities with antiferromagnets~\cite{optomagnonics_parvini} and magnetized epsilon-near-zero materials~\cite{optomagnonics_bittencourt_2, optomagnonics_bittencourt_3}.
Current work focuses on the design and optimization of optomagnonic systems on the nanoscale~\cite{OM-crystal, optomagnonic_new_1, optomagnonic_new_2, optomagnonics_haigh_5, optomagnonic_new_4}.
This is an important task to guide future experimental endeavors, since the state-of- the-art coupling strength is far below the predicted theoretical maximum value, mostly due to mode mismatch~\cite{OM-crystal, optomagnonics_haigh, optomagnonics_osada, optomagnonics_zhang, optomagnonics_osada_3, optomagnonics_haigh_2}. 

Due to the complexity of such geometries, the coupling needs to be calculated and optimized numerically. 
Most calculations however, involve simulation approaches which disregard several physical effects originating from the interaction of the light and the magnetization, e.g.~\cite{OM-crystal, OM_disk}.
In particular, magneto-optical effects~\cite{zvezdin_modern_1997} which stem from the splitting of the atomic energy levels due to the presence of a magnetic field, are not fully considered. 
A much more powerful approach would be to perform electromagnetic simulations which take the magnetic properties of the material into account.
Since the Finite-Difference Time-Domain (FDTD) method~\cite{yee_1966} is a broadly applicable and powerful numerical approach for computational electrodynamics, this method seems to be a suitable candidate for this goal.
Although such approaches have been already explored in the past to a certain extent~\cite{magnetic_dielectric_meep, magnetic_dieletric_fdtd_1, magnetic_dieletric_fdtd_2, magnetic_dieletric_fdtd_3}, none of them has been used to explore the effects of the magneto-optical interaction in detail.
Furthermore, these approaches cannot be used to perform the very computationally expensive simulations needed for the exploration of large 3D geometries with high resolution.
Thus, in this work we extend the in-house FDTD code~\cite{antonio_2015} developed for high performance nanoplasmonic computations to also treat magnetic dielectrics, by implementing the effective permittivity tensor modeling magneto-optical effects.
In particular, we explore the Faraday effect.

As a benchmark system, we investigate the plane wave scattering of light on a magnetized sphere in the Mie regime.
We choose this problem since it is analytically treatable as we show below, and has also already been investigated using theoretical predictions assisted with numerical tools~\cite{magnetic_sphere_1, magnetic_sphere_2, magnetic_sphere_3, magnetic_sphere_4, magnetic_sphere_5}.
As a material, we choose the ferrimagnet Yttrium-Iron-Garnet (YIG), the material of choice in optomagnonics due to its good magnetic and optical properties~\cite{yig_1, saga_of_yig}.

This work is structured as follows: In section (II) we give a brief introduction to magneto-optical effects and derive the corresponding effective permittivity tensor to be implemented into the FDTD method. Furthermore, we derive the theoretical expressions for the considered scattering problem using basic Mie theory~\cite{mie} extended to magnetic materials using the effective permittivity tensor. In section (III) we give a brief introduction to the FDTD technique and show how to implement the effective permittivity tensor to extend the FDTD method to treat also magnetic dielectrics. After discussing the simulation approach of our benchmark system in section (IV) we compare the theoretical predictions to the simulation outcome. 

\section{Mie light scattering of a magnetic sphere}
In the following we study the elastic scattering of optical photons impinging on a ferromagnetic sphere in the Mie regime, where the radius of the sphere is comparable to the photon's wavelength.
We consider an incident plane wave polarized along $\hat{\mathbf{x}}$ and propagating along $\hat{\mathbf{z}}$ with the electric field
\begin{equation}
    \mathbf{E}_{\text{in}} = \text{E}_0 \, e^{i k_0 z} \, \hat{\mathbf{x}}
\end{equation}
The sphere is assumed to be magnetized along $\hat{\mathbf{z}}$ with magnetization 
\begin{equation}
    \mathbf{M} = \text{M}_{\text{s}} \hat{\mathbf{z}}
\end{equation}
where $\text{M}_{\text{s}}$ is the saturation magnetization.
We note that this setting represents the so-called Faraday geometry (see Sec~\ref{sec:mo}).
The electromagnetic fields are governed by Maxwell's equations (assuming a time dependence $e^{-i \omega t}$)
\begin{equation}
    \begin{split}
        \nabla \cdot \mathbf{D} &= 0 \\[0.1cm]
        \nabla \cdot \mathbf{B} &= 0 \\[0.1cm]
        \nabla \times \mathbf{E} &= i \omega \mathbf{B} \\[0.1cm]
        \nabla \times \mathbf{H} &= - i \omega \mathbf{D}
    \end{split}
\end{equation}
with the magnetizing field $\mathbf{H} = \mathbf{B}/\mu_0$ where $\mu_0$ is the permeability of free space.
Outside the sphere, the displacement field is $\mathbf{D} = \mathbf{E}/\varepsilon_0$ with $\varepsilon_0$ the permittivity of free space. 
Inside the sphere, the permittivity is a magnetization dependent tensor $\mathbf{D}_i = \varepsilon_{ij} \mathbf{E}_j$ which is discussed below (see Sec.~\ref{sec:mo}).

Since the incident fields are produced by external sources, they are solutions of Maxwell's equations which are finite everywhere regardless of the presence of the sphere.
These solutions for a given frequency $\omega$ and wave vector $k_0=\omega/c$ are characterized by the angular indices $\{l,m\}$ and the polarization $\sigma\in\{\mathrm{TE},\mathrm{TM}\}$.
Thus, we can expand any incident field as (see Sup.~A.1)
\begin{equation}
    \boldsymbol{\mathbf{E}}_{\text{in}}(\boldsymbol{\mathbf{r}}) = \sum_{lm} \frac{1}{k_0r} \left[ \mathcal{E}_{lm}^{\text{in}} \boldsymbol{\mathbf{E}}_{lm,\text{TE}}^{\text{in}}(\boldsymbol{\mathbf{r}}) + c\mathcal{B}_{lm}^{\text{in}} \boldsymbol{\mathbf{E}}_{lm,\text{TM}}^{\text{in}}(\boldsymbol{\mathbf{r}}) \right], \label{Incident_expansion}
\end{equation}
where the coefficients $\{\mathcal{E}_{lm}^{\text{in}},\mathcal{B}_{lm}^{\text{in}} \}$ are the amplitude of TE and TM waves respectively. 
The eigenmodes are explicitly given by
\begin{equation}
\begin{split}
    \boldsymbol{\mathbf{E}}^{\text{in}}_{lm,\text{TE}}(\boldsymbol{\mathbf{r}}) &= S_l(k_0 r) \, \boldsymbol{\mathbf{Y}}^m_l(\theta, \phi), \\[0.1cm]
    \boldsymbol{\mathbf{E}}^{\text{in}}_{lm,\text{TM}}(\boldsymbol{\mathbf{r}}) &= -\sqrt{l(l+1)} \, \frac{S_l(k_0r)}{k_0 r} \, \boldsymbol{\mathbf{X}}^m_l(\theta, \phi) + i S_l^{'} (k_0 r) \, \boldsymbol{\mathbf{Z}}^m_l(\theta, \phi)
    \label{Defs:EMeigenmodes}
\end{split}
\end{equation}
with $S_l$ the Riccatti-Bessel functions (see Sup. Eq.~[6]) and $\mathbf{R}_l^m$ the vector spherical harmonics (see Sup. Eq.~[2]).
For a plane wave, the coefficients for the expansion turn out to be (see Sup.~A.2)
\begin{equation}
    \begin{split}
        \mathcal{E}^{\text{in}}_{lm} &= \text{E}_0 \, \sqrt{\pi (2l + 1)} \, i^l \, \left[ \delta_{m,1} + \delta_{m,-1} \right],\\[0.1cm]
        \mathcal{B}^{\text{in}}_{lm} &= \frac{\text{E}_0}{c} \, \sqrt{\pi (2l + 1)} \, i^{l+1} \, \left[ \delta_{m,1} - \delta_{m,-1} \right].
    \end{split}
\end{equation}

A similar expansion as in Eq.~\eqref{Incident_expansion} can be defined for scattered (outgoing) fields
\begin{equation}
    \boldsymbol{\mathbf{E}}_{\text{S}}(\boldsymbol{\mathbf{r}}) = \sum_{lm} \frac{1}{k_0r} \left[ \mathcal{E}_{lm}^{\text{S}} \boldsymbol{\mathbf{E}}_{lm,\text{TE}}^{\text{S}}(\boldsymbol{\mathbf{r}}) + c\mathcal{B}_{lm}^{\text{S}} \boldsymbol{\mathbf{E}}_{lm,\text{TM}}^{\text{S}}(\boldsymbol{\mathbf{r}}) \right], \label{scattered-fields}
\end{equation}
with expansion coefficients $\{\mathcal{E}^{{\text{S}}}_{lm},\mathcal{B}^{{\text{S}}}_{lm}\}$. Here, $\mathbf{E}_{lm,\sigma}^S$ are given by replacing $S_l\rightarrow\xi_l$ in Eqs.~\eqref{Defs:EMeigenmodes}, where $\xi_l$ are outgoing radial waves (see Sup. Eq.~[8]). In the remaining section, we derive the scattered coefficients in terms of the incident coefficients. 
 
\subsection{Magneto-optical effects}
\label{sec:mo}
In general, magneto-optical effects~\cite{zvezdin_modern_1997, magneto_optical_effects} refer to changes in the optical polarization states upon interaction with materials that are either subjected to an external magnetic field or magnetically ordered (or both).
In all cases, a magnetic field is present, either externally or internally, which causes a splitting of the atomic energy levels in the system with which the light can interact via electric-dipole transitions. 
This splitting, in general, creates different quantum states with non-degenerate energies and angular momenta.
As a consequence, the different polarization states of the light can interact differently with the material, since each polarization state interacts with a different non-degenerate energy level.
This leads to optical anisotropy which can be observed as birefringence.

In general, the splitting of the energy levels due to an external field is caused by two main mechanisms: The Zeeman effect~\cite{zeeman_splitting} which refers to the energy splitting in the presence of an external field and the spin-orbit coupling~\cite{light_scattering_on_magnons_1} which refers to the splitting due to the spin-orbit interaction.
Which effect is causing the splitting, highly depends on the material. 
In total, there are two classes of materials and a third representing a transition between the two~\cite{pershan}: (i) Diamagnetic, transparent solids with at least uniaxial symmetry where magneto-optical effects can only be caused by the Zeeman effect. (ii) Paramagnetic and ferromagnetically ordered materials where magneto-optical effects are predominately caused by the spin-orbit coupling which usually exceeds the Zeeman interaction. (iii) Semiconductors and non-ferromagnetic metals represent a transition between the two. Usually, both effects are present and non-negligible.

Considering only the effects affecting the polarization of the transmitted light, magneto-optical effects can be classified into two classes according to the relative orientation of the light wave vector $\boldsymbol{\mathbf{k}}$ with respect to the magnetic field $\boldsymbol{\mathbf{H}}$: (i) The Faraday geometry with $\boldsymbol{\mathbf{k}} \parallel \boldsymbol{\mathbf{H}}$  and (ii) the Voigt geometry with $\boldsymbol{\mathbf{k}} \perp \boldsymbol{\mathbf{H}}$.
In the case of the Faraday geometry, birefringence occurs since the two circularly polarized components of the light effectively see different refractive indices resulting in the so-called magnetic circular birefringence or the Faraday effect.
In the case of the Voigt geometry the two linearly polarized components see different refractive indices, causing the so-called magnetic linear birefringence or the Cotton-Mouton/ Voigt effect.
 
In this work, we focus on the Faraday effect only.
As discussed above, due to the magnetic fields present, phenomenologically the two circular polarized components of the linear polarized light see different refractive indices $n_+$ and $n_-$ what causes them to propagate with different speeds, $c/n_+$ and $c/n_-$, through the medium.
As a consequence, the two polarization states acquire a phase shift resulting in an overall rotation of a linearly polarized light.
The angle of rotation is called the Faraday rotation $\theta_{\text{F}}$ which can be expressed as~\cite{zvezdin_modern_1997}
\begin{equation}
    \theta_{\text{F}} = \frac{\omega}{2c} (n_+ - n_-)L
\end{equation}
with $\omega$ the frequency of the light, $c$ the speed of light, and $L$ the propagation length of the light.

For modeling magneto-optical effects we need to introduce an effective permittivity tensor.
We note that in general, magneto-optical effects can be entirely modeled by either using the effective permittivity or the effective permeability~\cite{pershan}.
However, when working at optical frequencies it is difficult to give clear physical interpretation to the magnetization and thus the permeability~\cite{spin_waves}. 
As a consequence, it is common to use the permittivity only and set the permeability to that of vacuum.

In the case of dispersive magnetic materials, the modified effective permittivity due to the Faraday effect is given by (see Sup.~B)
\begin{equation}
\overline{\varepsilon}(\mathbf{M}, \omega) =
\varepsilon_{0}
\begin{bmatrix}
\varepsilon_r(\omega) & - i f_{{F}} (\omega) M_z &  i f_{\text{F}} (\omega) M_y\\
 i f_{{F}} (\omega) M_z & \varepsilon_r(\omega) & - i f_{{F}} (\omega) M_x\\
 - i f_{{F}} (\omega) M_y &  i f_{{F}} (\omega) M_x & \varepsilon_r(\omega)
\end{bmatrix},
\label{eq:eps_mag}
\end{equation}
where $\varepsilon_0$ is the vacuum permittivity, $\varepsilon_r$ the relative permittivity, and $f_{\text{F}}$ is a material dependent constant related to the Faraday rotation via $f_{\text{F}} = [(2c\sqrt{\varepsilon_r})/(\omega \text{M}_{\text{s}})] \theta_{\text{F}}$.

For modeling $\varepsilon_r(\omega)$ and $f_{\text{F}}(\omega)$ we consider a specific dispersion model~\cite{optomagnonics_bittencourt_2}.
Since in magnetic dielectrics, the Faraday effect microscopically originates from electric dipole transitions between degenerate ground and excited states due to the energy splitting e.g. caused by the spin-orbit coupling and the Zeeman interaction~\cite{optomagnonics_bittencourt_2, pershan, light_scattering_on_magnons_1}, our minimal model considers a single-resonance Lorentz-like dispersion model~\cite{optomagnonics_bittencourt_2}
\begin{equation}
    \begin{split}
        \varepsilon_r(\omega) &= 1 + \frac{\omega_0^2 (\varepsilon_r - 1)}{\omega_0^2 - \omega^2 - i \eta \omega},\\[0.2cm]
        f_{\text{F}}(\omega) &= \frac{A_3 \omega \omega_0}{(\omega_0^2 - \omega^2 - i \eta \omega)^2},
    \end{split}
    \label{eq:toy-model}
\end{equation}
where $\omega_0$ is the resonance frequency for the ionic transitions in the absence of perturbations, $\varepsilon$ the permittivity for $\omega << \omega_0$, $\eta$ an absorption factor (required by causality), and $A_3$ a quantity which depends on the spin-orbit coupling and the electric dipole matrix.
We note that this model assumes zero orbital angular momentum of the ground state.
Although this model can only be seen as a theoretical toy model since it disregards several resonances of the material, it nonetheless represents an appropriate model for our purposes~\cite{optomagnonics_bittencourt_2}.

\subsection{Scattering coefficients of the scattered fields}
\label{sec:scattering_coeffs}
In the following we consider YIG as the basis material,  for which the Faraday rotation per unit length $\theta_{\text{F}}$ is smaller than $10^4 \si{\per\meter}$~\cite{yig_2}. 
Inside a  sphere with radius $\SI{1}{\micro\meter}$, we expect a rotation of $<0.01$, implying that we can treat the Faraday effect as a perturbation.
Thus, up to linear order in $f_{\text{F}}$, the expansion coefficients in Eq.~\eqref{scattered-fields} are 
\begin{equation}
    \mathcal{E}^{\rm S}_{lm} = \mathcal{E}^{\rm M}_{lm} + \mathcal{E}^{\rm F}_{lm},
\end{equation}
where the (unperturbed) Mie scattered light coefficients are (see Sup.~A.3)
\begin{equation}
    \begin{split}
        \mathcal{E}^{\text{M}}_{lm}(k_0) &= \text{E}_0 \, r_l^{\text{E}}(k_0) \, \sqrt{\pi (2l + 1)} \, i^l \, \left[ \delta_{m,1} + \delta_{m,-1} \right]\\[0.1cm]
        \mathcal{B}^{\text{M}}_{lm} (k_0) &= \frac{\text{E}_0}{c} \, r_l^{\text{B}}(k_0) \, \sqrt{\pi (2l + 1)} \, i^{l+1} \, \left[ \delta_{m,1} - \delta_{m,-1} \right]
    \end{split}
\end{equation}
and the expansion coefficients of the Faraday scattered fields are (see Sup.~A.4)
\begin{equation}
    \begin{split}
        \mathcal{E}^{\text{F}}_{lm}(k_0) &= f \text{E}_0 \sqrt{\pi} \, \left[ P^{\text{E}}_{l-1,1}(k_0) \sqrt{2l-1} - M^{\text{E}}_{l+1,1}(k_0) \sqrt{2l+3} \right] \, i^{l} \, \left[ \delta_{m,1} - \delta_{m,-1} \right]\\[0.1cm]
        \mathcal{B}^{\text{F}}_{lm}(k_0) &= f \frac{\text{E}_0 \sqrt{\pi}}{c}  \, \left[ P^{\text{B}}_{l-1,1}(k_0) \sqrt{2l-1} - M^{\text{B}}_{l+1,1}(k_0) \sqrt{2l+3} \right] \, i^{l-1} \, \left[ \delta_{m,1} + \delta_{m,-1} \right]
    \end{split}
    \label{farad_coeff_l_combi}
\end{equation}
The expressions for $r_l^{\text{E}}$, $r_l^{\text{B}}$, $P^{\text{E}}_{lm}$, $P^{\text{B}}_{lm}$, $M^E_{lm}$, and $M^B_{lm}$ can be found in Sup.~A.3 and~A.4.
 
\begin{figure}[t!]
\begin{centering}
\includegraphics[width=1\columnwidth]{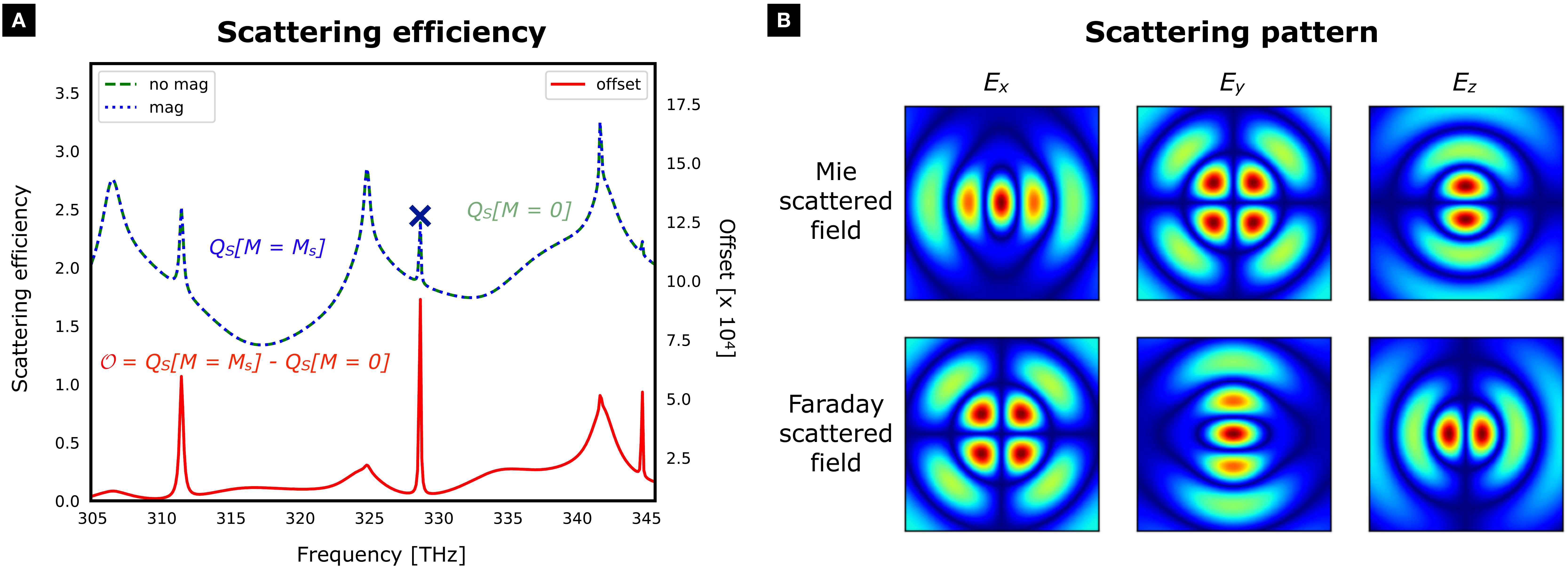}
\par\end{centering}
\caption{Light scattering on a magnetic sphere with radius $R=\SI{1}{\micro\metre}$ using Mie theory: (A) Scattering efficiency of a magnetic (blue) and non-magnetic sphere (green) in the frequency range $\SI{305}{\THz}$ to $\SI{345}{\THz}$. The peaks of the offset $\mathcal{O} = Q_{\text{S}}[\mathbf{M} = \text{M}_{\text{s}}] - Q_{\text{S}}[\mathbf{M} = 0]$ indicate that the scattering efficiency of the magnetized sphere is broader at the peaks than the scattering efficiency of the un-magnetized sphere which is a result of the energy level splitting of the modes in the presence of a magnetic field. (B) Electric field pattern of the scattered field with frequency $\SI{329}{\tera \hertz}$ (wavelength $\SI{912.95}{\nano\meter}$, identified as TE-10, see cross in A) in a plane perpendicular to the propagation direction of the incident wave ($xy$-plane) in the far field using a Cartesian basis (see Fig.~\ref{fig_sim_domain}B dashed line, $z \sim \SI{3}{\micro \metre}$ measured from the center of the sphere in propagation direction of the incoming wave). The Faraday scattered field components are rotated by $90^{\circ}$ compared to the Mie scattered field components and $E_x$ and $E_y$ interchanged patterns which is a direct result of the Faraday rotation.\\ Note: In A and B we used the Lorentz model in Eq.~\eqref{eq:toy-model} for $f_{\text{F}}$ (average value in the investigated frequency: $f_{\text{F}} =0.0002$).}
\label{fig_theory_pred}
\end{figure}

Using the azimuthal dependence of vector spherical harmonics $\mathbf{R}_l^m \propto e^{im\phi}$, we can deduce the $\phi$ dependence of both scattered fields,
\begin{equation}
\begin{split}
    \mathbf{E}_{\text{M}} &\propto \cos \phi \\[0.1cm]
    \mathbf{E}_{\text{F}} &\propto \sin \phi,
\end{split}
\end{equation}
showing that the Mie and the Faraday scattered field are azimuthally orthogonal. 
In a Cartesian basis this means, as we see in Fig.~\ref{fig_theory_pred}B, that all field components of the Faraday scattered field are rotated by $90^{\circ}$ compared to the field components of the Mie scattered field and that $E_x$ and $E_y$ interchange patterns. 
This is a result of the polarization rotation through the Faraday effect.

\subsection{Scattering efficiency}
\label{sec:sacttering_efficiency}
For deriving the scattering cross-section and the scattering efficiency we also expand the magnetic field into vector spherical harmonics by using the following substitutions
\begin{equation}
    \begin{split}
    \mathcal{E}_{lm}^{\text{i}} &\rightarrow \mathcal{B}_{lm}^{\text{i}} \\[0.1cm]
    c\mathcal{B}_{lm}^{\text{i}} &\rightarrow \frac{\mathcal{E}_{lm}^{\text{i}}}{c}
    \end{split}
    \label{eq:conv-e-b}
\end{equation}
with ${\text{i}} = [\text{in}, \text{{M}}, \text{{F}}]$.
In general, the cross-section can be expressed as~\cite{bohren_and_huffman}
\begin{equation}
    \sigma_{\text{S}} = \frac{W_{\text{S}}}{I_{\text{in}}}
\end{equation}
with $W_{\text{S}}$ the energy flux of the scattered light through a certain surface and $I_{\text{in}}$ the intensity of the incoming plane wave.
In spherical coordinates the energy flux can be found via~\cite{bohren_and_huffman}
\begin{equation}
    W_{\text{S}} = \frac{1}{2} \, \mathcal{R}e \left[ \int r^2 \sin{\theta} \, d\theta d\phi \, \left( E_{{\text{S}}\,\theta} H_{{\text{S}} \, \phi}^* - E_{{\text{S}}\,\phi} H_{{\text{S}} \, \theta}^* \right)\right]
\end{equation}
with $\boldsymbol{\mathbf{E}}_{\text{S}} = \boldsymbol{\mathbf{E}}_{\text{M}} + \boldsymbol{\mathbf{E}}_{\text{F}}$ and $\boldsymbol{\mathbf{H}}_{\text{S}} = \boldsymbol{\mathbf{B}}_{\text{S}}/\mu_0 = (\boldsymbol{\mathbf{B}}_{\text{M}} + \boldsymbol{\mathbf{B}}_{\text{F}})/\mu_0$ the total scattered fields.
Since the incoming wave is a plane wave, its intensity is given by
\begin{equation}
    I_{\text{in}} = \frac{\text{E}_0^2}{2 \eta}
\end{equation}
with the impedance $\eta = \sqrt{\mu_0/{n^2 \varepsilon_0}}$.
Dividing the scattering cross-section by the geometric cross-section of the sphere gives the scattering efficiency
\begin{equation}
    Q_{\text{S}} = \frac{\sigma_{\text{S}}}{\sigma_{\text{geom}}} = \frac{\sigma_{\text{S}}}{\pi R^2}.
    \label{eq:Q_S_theory}
\end{equation}
Fig.~\ref{fig_theory_pred}A shows the scattering efficiencies for a magnetized and an un-magnetized sphere in the frequency range $\SI{305}{\THz}$ to $\SI{345}{\THz}$. 
For ``measuring" their difference, we introduce the following offset measure
\begin{equation}
    \mathcal{O} = Q_{\text{S}}[\mathbf{M} = \text{M}_{\text{s}}] - Q_{\text{S}}[\mathbf{M} = 0]
    \label{eq:offset}
\end{equation}
which is also plotted in Fig.~\ref{fig_theory_pred}A.
As the peaks of the offset only occur at the scattering peaks, we can deduce that the scattering efficiency of the magnetized sphere is broader at the peaks than the scattering efficiency of the un-magnetized sphere.
This is a result of the energy mode splitting.
We note that in contrast to~\cite{magnetic_sphere_1} we do not observe a splitting of the peaks since we only consider the effects that are linear in $f_{\text{F}}$ whereas the splitting is a second order effect. This is valid because the typical line widths of Mie resonances are larger than the small splittings.


\section{FDTD approach for magneto-optical effects}
Below we show how the FDTD method can be extended to treat magnetic dispersive media and thus to incorporate magneto-optical effects.

\subsection{The FDTD method}
The FDTD technique is a numerical method to perform electromagnetic simulations.
The success of this method is based on its simplicity combined with its broad applicability especially to many complex materials in arbitrary shape configurations and for broad bandwidths.
The FDTD method belongs to the finite difference methods solving the time-dependent Maxwell's equations~\cite{fdtd_workshop}
\begin{equation}
    \begin{split}
        \frac{\partial \boldsymbol{\mathbf{D}} (\boldsymbol{\mathbf{r}}, t)}{\partial t} &= \nabla \times \boldsymbol{\mathbf{H}} (\boldsymbol{\mathbf{r}}, t) - \mathbf{J}_{{free}} (\boldsymbol{\mathbf{r}}, t), \\[0.1cm]
        \frac{\partial \boldsymbol{\mathbf{B}} (\boldsymbol{\mathbf{r}}, t)}{\partial t} &= -\nabla \times \boldsymbol{\mathbf{E}} (\boldsymbol{\mathbf{r}}, t),
    \end{split}
    \label{eq:maxwell}
\end{equation}
alongside $\boldsymbol{\mathbf{D}} (\boldsymbol{\mathbf{r}}, t) = \varepsilon_0 \boldsymbol{\mathbf{E}} (\boldsymbol{\mathbf{r}}, t) + \boldsymbol{\mathbf{P}} (\boldsymbol{\mathbf{r}}, t)$ and $\boldsymbol{\mathbf{H}} (\boldsymbol{\mathbf{r}}, t) = 1/\mu_0 \boldsymbol{\mathbf{B}} (\boldsymbol{\mathbf{r}}, t) + \boldsymbol{\mathbf{M}} (\boldsymbol{\mathbf{r}}, t)$. 
These coupled differential equations are transformed into numerical equations by using the central differencing scheme
\begin{equation}
    \frac{df}{dx} = \frac{f(x + h/2) - f(x - h/2)}{h}.
\end{equation}
Following the so-called Yee algorithm\cite{yee_1966}, time and space derivatives are substituted by their corresponding central differencing formulas and the simulation domain is discretized with a regular structured rectangular grid. 
For maintaining second order accuracy of the central differencing operators, the Yee algorithm staggers the electric and magnetic fields in both time and space.
This means that in time Maxwell's equations are solved in a leapfrog manner, meaning that first the electric field vector components are solved in a given space region at a given instant of time, then the magnetic field vector components are solved in the same region at the next instant of time.
Besides the staggering in time, also the spatial grids of both fields are staggered meaning that the electric field vector components are located midway between a pair of magnetic field vector components.
These staggered grids result in the typical FDTD relation: At any given point in space, the updated E-field (H-field) in time depends on the stored value of the E-field (H-field) and the numerical curl of the surrounding H-field (E-field) in space.
This nicely represents the given relations in Maxwell's equation where the change of the E-field (H-field) in time is dependent on the spatial change in the H-field (E-field)~\cite{yee_1966, intro_to_fdtd, understanding_fdtd, fdtd_workshop}.

One big advantage of the FDTD method is that almost any dispersive material can be modeled.
Since dispersive media exhibit a frequency dependent susceptibility $\chi(\omega)$, the expression for the susceptibility needs to be known to derive the polarization vector ~\cite{fdtd_workshop}
\begin{equation}
    \boldsymbol{\mathbf{P}}(\omega) = \varepsilon_0 \chi(\omega) \boldsymbol{\mathbf{E}}(\omega)
    \label{eq:pol}
\end{equation}
which then includes the dispersive material properties via the update equations for the electric field. Thus, for implementing a dispersive material using a model for $\chi(\omega)$ we need to introduce an auxiliary differential equation in the leap frog scheme for the polarization vector $\boldsymbol{\mathbf{P}}$~\cite{fdtd_workshop}.

\subsection{Extension to dispersive magnetic materials}
As already mentioned, for implementing a material into the FDTD method the material's susceptibility tensor needs to be known as a function of frequency to adapt the update equations.
Since the susceptibility is related to the permittivity via $\chi = \varepsilon - 1$, the frequency dependent susceptibility of a magnetic material can be derived by using the frequency dependent permittivity tensor in Eq.~\eqref{eq:eps_mag} in combination with the introduced single-resonance model in Eq.~\eqref{eq:toy-model}. 

For implementing the susceptibility model into the update equations of the electric field we need the polarization vector $\boldsymbol{\mathbf{P}}$ which in our case is given by
\begin{equation}
        \boldsymbol{\mathbf{P}} = \varepsilon_0 \, \frac{\omega_0^2 (\varepsilon_r - 1)}{\omega_0^2 - \omega^2 - i \eta \omega} \, \boldsymbol{\mathbf{E}} 
        + i \omega \varepsilon_0 \, \frac{A_3 \omega_0}{(\omega_0^2 - \omega^2 - i \eta \omega)^2} \, \left[ \boldsymbol{\boldsymbol{\mathbf{M}}} \times \boldsymbol{\mathbf{E}} \right].
\end{equation}
This equation describes the optical response of a magnetic material in the frequency domain, for transforming it into the time domain we need to apply a Fourier transform (see Sup.~C.1) leading to
\begin{equation}
    \begin{split}
        \partial_t^4 \boldsymbol{\mathbf{P}} + 2 \eta \, \partial_t^3 \boldsymbol{\mathbf{P}} + &(2 \omega_0^2 + \eta^2) \, \partial_t^2 \boldsymbol{\mathbf{P}} + 2 \omega_0^2 \eta \, \partial_t \boldsymbol{\mathbf{P}} + \omega_0^4 \boldsymbol{\mathbf{P}} \\
        &= \varepsilon_0 \omega_0^2 (\varepsilon - 1) \left[ \partial_t^2 \boldsymbol{\mathbf{E}} + \eta \, \partial_t \boldsymbol{\mathbf{E}} + \omega_0^2 \boldsymbol{\mathbf{E}} \right] - \varepsilon_0 A_3 \omega_0 \, \left[ \boldsymbol{\boldsymbol{\mathbf{M}}} \times \partial_t \boldsymbol{\mathbf{E}} \right].
    \end{split}
\end{equation}
We note that we assumed a slowly varying magnetization $\boldsymbol{\boldsymbol{\mathbf{M}}}$ in time compared to the electric field $\boldsymbol{\mathbf{E}}$, thus $\partial_t \left[ \boldsymbol{\boldsymbol{\mathbf{M}}} \times \boldsymbol{\mathbf{E}} \right] = \left[ \boldsymbol{\boldsymbol{\mathbf{M}}} \times \partial_t \boldsymbol{\mathbf{E}} \right]$.
The update equations for the polarization vector are then obtained by using central differencing (see Sup.~C.2) which then includes the magnetic model via the update equations for the electric field in the FDTD method.

\subsection{Model for YIG}
In order to benchmark our numerical method, we consider the material  Yttrium-Iron-Garnet, a typical choice for magneto-optical phenomena.  For adapting our model to the material parameters of YIG, we need to find the values for the unknown parameters ${\varepsilon}$, $\eta$, and $A_3$.
Although YIG has several absorption lines between $\SI{400}{\nano\metre}$ and $\SI{900}{\nano\metre}$ (see~\cite{yig_1} Fig. 1), the broad and strong absorption line at $\sim \SI{500}{\nano\metre}$ gives a strong contribution to the Faraday effect (see~\cite{yig_2} Fig. 2). We therefore set the resonance frequency to $\omega_0= 2\pi c/\lambda = 2\pi \times \SI{600}{\tera\hertz}$ which sets the parameters to ${\varepsilon}_r = 4.9$ and $A_3 \sim -2.25 \cdot 10^{22}\, \text{rad}^2\text{Hz}^2\text{m}/\text{A}$~\cite{yig_3}.
Since we are working with a single-resonance model, the absorption parameter $\eta$ can be set to the absorption coefficient of YIG $\eta/\omega_0 \sim 10^{-6}$ (@ $\SI{1.2}{\micro\metre}$)~\cite{victor_1}. Note that we use a single frequency $\omega_0$ for benchmarking purposes. For a realistic model of the scattering, all absorption lines within the frequency range of interest should be considered.


\section{Benchmarking the extended method}
Following the analytical predictions derived in Secs~\ref{sec:scattering_coeffs} and~\ref{sec:sacttering_efficiency}, we simulate the scattering of a plane wave which is propagating along $\hat{\mathbf{z}}$ and polarized along $\hat{\mathbf{y}}$ on a magnetic YIG sphere with radius $R = \SI{1}{\micro \metre}$ fully magnetized along $\hat{\mathbf{z}}$.
Since we want to reconstruct the features stemming from the Faraday effect derived in the theory discussion, our aim is to obtain the scattering efficiency and the scattered mode profile with the extended FDTD code.

\subsection{Simulation approach}
\begin{figure}[t!]
\begin{centering}
\includegraphics[width=1\columnwidth]{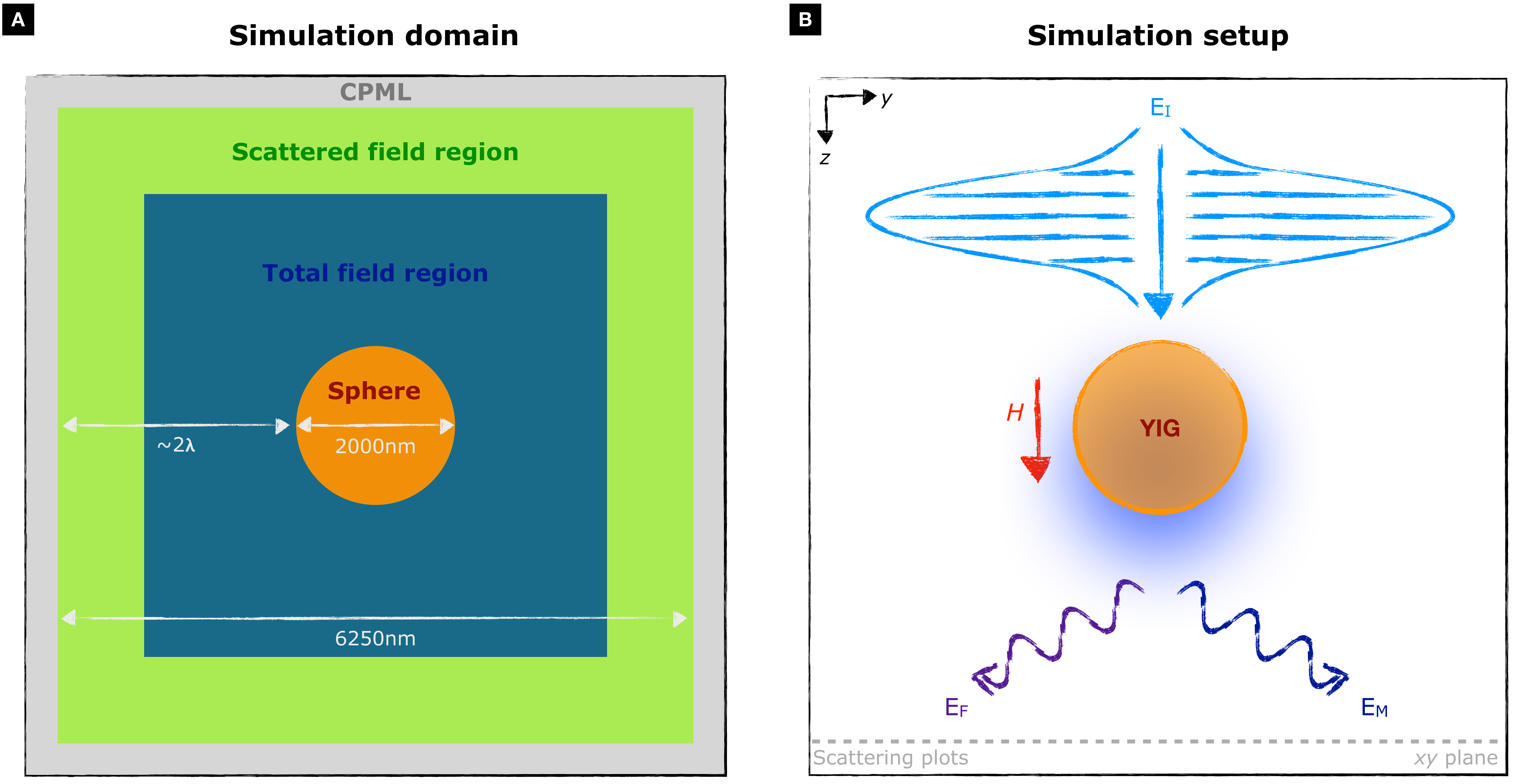}
\par\end{centering}
\caption{Simulation domain and setup: (A) Simulation domain consisting of (i) the YIG sphere (red), (ii) the total field region (blue) containing all simulated fields (iii) the scattered field region (green) containing only the scattered fields, and (iv) the convolutional perfectly matched layers (CPML) region (gray) truncating the computational domain to a finite volume (note that the actual simulation domain is 3D). (B) Sketch of the simulation setup. The incident field in the form of a Gaussian plane wave pulse propagates along the magnetization direction of the sphere and is scattered into the Mie and Faraday scattered part. The dashed line indicates the plane the spatial mode shapes were evaluated in Figs.~\ref{fig_theory_pred} and~\ref{fig_benchmark}.}
\label{fig_sim_domain}
\end{figure}

Since we are simulating a scattering problem, beside the sphere we also need to simulate the air surrounding the sphere which needs to be truncated to a finite domain.
Thus, we choose a cubic simulation domain which is subdivided into four regions (see Fig.~\ref{fig_sim_domain}A): (i) the YIG sphere, (ii) the total field region surrounding the sphere and containing all simulated fields, the incident and the scattered fields, (iii) the scattered field region which contains only the scattered fields, and (iv) the convolutional perfectly matched layers (CPML) region which truncate the computational domain to a finite volume but simulating free space by absorbing all outgoing waves.
Since the near fields should have decayed before reaching the CPML region, the distance between the sphere surface and the CPML layers should be large enough, ideally $\sim 3 \lambda$.
Due to computational constraints, we however chose a distance of $\sim 2 \lambda$ which should still account for a sufficient decay of the near fields.
Assuming the wavelength of the scattered field to be $\sim \SI{1000}{\nm}$, our simulation domain has a side length of $D =  \SI{6250}{\nm} \sim 2 (R+2\lambda) $ (excluding the CPML region). 
For good absorbing effects at the air boundaries the CPML layer is 30 cells thick (regardless of the actual cell size).
Furthermore, the total field - scattered field boundary is set to be 40 cells away from the CPML layer (again regardless of the actual cell size).

For discretizing the simulation domain, we use an uniform per-component staircasing grid which assigns the permittivity for each electric field component based on the position of the sampling point in the simulation cell relative to the sphere.
This means if the sampling point lies inside the sphere, the permittivity of YIG is assigned, if the point is outside the sphere the permittivity of air is assigned.
This is in contrast to a staircase grid where the permittivity is assigned to the whole cell~\cite{antonio_2015}.
For avoiding numerical dispersion, we choose a uniform cell size of $\Delta d = \SI{10}{\nm} \sim \lambda/100$ giving $625 \times 625 \times 625$ cells to simulate.
Using the Courant stability condition~\cite{fdtd_workshop}
\begin{equation}
    \Delta t \leq \frac{1}{c \sqrt{1/ \Delta x^2 + 1/ \Delta y^2 + 1/ \Delta z^2}}
    \label{eq:courant}
\end{equation}
with $\Delta x = \Delta y = \Delta z = \Delta d$ the time step is set to $\mathit{\Delta} t = (\mathit{\Delta} d)/(\sqrt{3} c) = \SI{0.2}{\femto \second}$.

For exciting the system and for stimulating scattering we use a electromagnetic plane wave in the form of a Gaussian source
\begin{equation}
    \mathcal{G}(t) = \exp \bigg(- \frac{t^2}{\sigma^2}\bigg) \cos \big(2 \pi f_0 t\big)
\end{equation}
with width $\sigma = \SI{1.97e-11}{\s}$ and center frequency $f_0 = \SI{329}{\THz}$ ($\lambda_0 = \SI{912.95}{\nm}$, frequency of the mode of interest found by ``pre"-simulations).

\subsection{Benchmark}
For benchmarking the code against the theory predictions shown in Fig.~\ref{fig_theory_pred} we perform simulations for obtaining the scattering efficiency and the spatial field patterns of the scattered fields.

\subsubsection{Scattering efficiency}
We start with the scattering efficiency which we need to evaluate for a magnetized ($Q_{\text{S}}[\mathcal{M} = \text{M}_{\text{s}}]$) and an un-magnetized sphere ($Q_{\text{S}}[\mathcal{M} = 0]$) implemented as~\cite{antonio_2015}
\begin{equation}
    Q_{\text{S}} = \frac{ \frac{1}{2}\int_S dS ~ \text{Re} \left[ \boldsymbol{\mathbf{E}} \times \boldsymbol{\mathbf{H}}^* \right] \cdot \hat{\mathbf{n}}}{ \frac{1}{2} \text{Re} \left[ \boldsymbol{\mathbf{E}}_{\text{in}} \times \boldsymbol{\mathbf{H}}_{\text{in}}^* \right] \, A_{\text{geom}}}
\end{equation}
being equal to the theory expression in Eq.~\eqref{eq:Q_S_theory}.
For both simulations, we evolve the system for $600000$ time steps and evaluate the scattering efficiency in the frequency window
$\SI{305}{\THz}$ to $\SI{346}{\THz}$ ($\SI{868}{\nm}$ to $\SI{984}{\nm}$) for $401$ points.

\begin{figure}[t!]
\begin{centering}
\includegraphics[width=1\columnwidth]{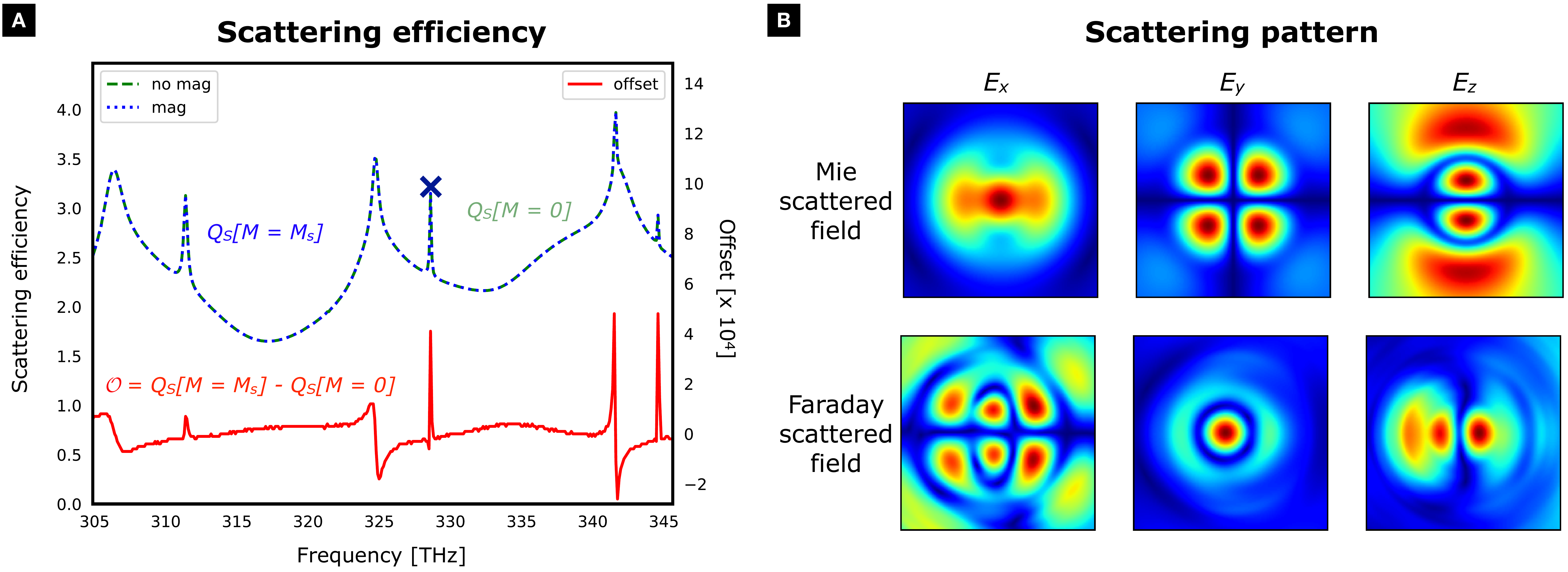}
\par\end{centering}
\caption{Light scattering on a magnetic sphere with radius $R=\SI{1}{\micro\metre}$ using the adapted FDTD code: (A) Scattering efficiency of a magnetic (blue) and non-magnetic sphere (green) in the frequency range $\SI{305}{\THz}$ to $\SI{345}{\THz}$. The peaks of the offset $\mathcal{O} = Q_{\text{S}}[\mathbf{M} = \text{M}_{\text{s}}] - Q_{\text{S}}[\mathbf{M} = 0]$ indicate that the scattering efficiency of the magnetized sphere is broader at the peaks than the scattering efficiency of the un-magnetized sphere which is a result of the splitting of the modes in the presence of a magnetic field. The reason for less pronounced peaks and negative contributions are precision issues in the simulation (B) Electric field pattern of the scattered field with frequency $\SI{329}{\tera \hertz}$ (wavelength $\SI{912.95}{\nano\meter}$, identified as TE-10, see cross in A) in a plane perpendicular to the propagation direction of the incident wave ($xy$-plane) in the far field using a Cartesian basis (see Fig.~\ref{fig_sim_domain}B dashed line, $z \sim \SI{3}{\micro \metre}$ measured from the center of the sphere in propagation direction of the incoming wave). The Faraday scattered field components are rotated by $90^{\circ}$ compared to the Mie scattered field components and $E_x$ and $E_y$ interchanged patterns which is a direct result of the Faraday rotation.}
\label{fig_benchmark}
\end{figure}

Fig.~\ref{fig_benchmark}A shows the two obtained scattering efficiencies and their offset defined in Eq.~\eqref{eq:offset}.
As we see, the scattering peaks coincide very well with the theoretical prediction shown in Fig.~\ref{fig_theory_pred}A and also have the same order of magnitude.
Also in the simulation, the peaks of the offset mostly occur at the scattering peaks, although they are much less pronounced and more noisy compared to the theory prediction.
The reason therefore might be that the contribution of the Faraday effect to the scattering is tiny and the precision in the simulation is not large enough. 
Furthermore, the offset shows negative components which might stem from a slight peak shift in the case of the magnetized sphere compared to the un-magnetized sphere.
Again, this could be due to a precision problem.
For overcoming this issue, a much finer mesh is necessary, however our computational power and time are limited such that we cannot use a much finer mesh.
Nonetheless, we can conclude that we observe a peak broadening in case of a magnetized sphere which is a result of the code taking the energy mode splitting into account. Since YIG has a small Faraday constant, the linewidth of the Mie resonances masks the magnetization-induced mode splitting, compare e.g. with Ref.~\cite{magnetic_sphere_1} where the splitting is resolved due to a much larger Faraday rotation constant of the considered material. 

\subsubsection{Scattered field patterns}
For simulating the scattered field patterns we also need to perform two simulations, one for the  magnetized ($\boldsymbol{\mathbf{M}} = \text{M}_{\text{s}} \hat{z}$) and one for the un-magnetized sphere ($\boldsymbol{\mathbf{M}} = 0$).
The simulation in the un-magnetized case then gives the Mie scattered fields $\boldsymbol{\mathbf{E}}_{\text{M}} = \boldsymbol{\mathbf{E}}[\boldsymbol{\mathbf{M}} = 0]$ and the Faraday scattered fields are deduced from both simulations via $\boldsymbol{\mathbf{E}}_{\text{F}} = \boldsymbol{\mathbf{E}}[\boldsymbol{\mathbf{M}} = \text{M}_{\text{s}} \hat{z}] - \boldsymbol{\mathbf{E}}[\boldsymbol{\mathbf{M}} = 0]$.
For both simulations, we evolve the system for $60000$ time steps and evaluate the scattering pattern for the frequency $f_0 = \SI{329}{\THz}$ ($\lambda_0 = \SI{912.95}{\nm}$), which is the mode of interest (see Figs.~\ref{fig_theory_pred}A and~\ref{fig_benchmark}A).
The actual scattered field patterns are obtained by monitoring the time evolution of the field in a plane perpendicular to the propagation axis of the incident plane wave located after the sphere (see Fig.~\ref{fig_sim_domain}B dashed line) and performing a running discrete Fourier transformation on the time-dependent fields in this plane.
The plane is placed in the 10th last cell layer in air before the CPML layer to monitor the furthest scattered part of the light.
We note that for these simulations we use less time steps compared to the simulations performed for the scattering efficiencies since the weak Faraday scattered fields are ``hidden" behind the stronger Mie scattering fields which are much longer present.
Furthermore, the FDTD method accumulates a larger error for longer evolution times which also might hide the small contribution of the Faraday scattered field part.

Fig.~\ref{fig_benchmark}B shows the obtained scattering profiles for the Mie and Faraday scattered field components for the desired scattered mode. 
As we see, the Mie scattered fields almost resemble the shape of the theory prediction shown in Fig.~\ref{fig_theory_pred}B, however their shape seems to be more ``zoomed" in.
The reason therefore might be that the theory gives the real far fields whereas the simulation outcome shows the near and intermediate fields since the simulation domain is truncated.
Although the obtained Faraday scattered fields are much more noisy, they still show the expected behavior: The $E_z$ component is rotated by $90^\circ$ and the $E_x$ and $E_y$ component interchanged shape.
The reason for the noisy shape for the $E_x$ component might be that this component has the smallest contribution in absolute value and thus the actual shape might be hidden behind noise.
For overcoming this issue, we also need to use a finer mesh and to increase the simulation domain such that we can evaluate the patterns in the real far field. Furthermore, a longer simulation time is required to resolve the sharp resonances.
As already mentioned, this is limited by the available computational power and time constrains.
Nonetheless, we can conclude that the code correctly obtains the rotation stemming from the Faraday rotation.


\section{Conclusion}
We proposed an approach to extend the FDTD method to treat magnetic dispersive media by using a modified effective permittivity encapsulating the Faraday effect.
As a benchmark system to test the functionality of the extended method, we considered the scattering of a plane wave on a fully magnetized YIG sphere in the Mie regime.
We theoretically showed that including the modified effective permittivity in Maxwell's equations results in broadened peaks in the scattering efficiency nicely showing the splitting of atomic energy levels when a magnetic field is present. 
Furthermore, the scattered fields include an additional part rotated with respect to the usual Mie scattered fields stemming from the light-matter interaction through the Faraday effect

We showed that the extended FDTD method can reproduce both predicted features: the peak broadening in the scattering efficiency and the rotation in the scattered fields.
These features are not as pronounced as in the analytical model due to the high Q of the considered modes, which are hard to resolve within FDTD.
For better results a higher resolution in space and longer simulation times are necessary, which is not feasible with our current computational resources.
For benchmarking the system further, other frequency ranges where the Faraday effect is more pronounced could be explored.
However, higher frequencies have even higher Q and thus, are harder to simulate using the FDTD method.

A possible field of application for the extended method might be magnetoplasmonics where the interaction of plasmonics and magneto-optical effects is explored.
In this context, an Epsilon-Near-Zero (ENZ) optomagnonic structure, as proposed e.g. in~\cite{optomagnonics_bittencourt_2}, could be investigated which serves as an additional route to enhance the optomagnonic coupling.
Furthermore, backaction in the optomagnonic coupling could be explored. 


\section{Acknowledgments}
We thank Victor Bittencourt and Jesse Thompson for insightful discussions. 
This research was performed as part of a collaboration within the Max Planck-University of Ottawa Centre for Extreme and Quantum Photonics, whose support all authors gratefully acknowledge.
J.G. acknowledges financial support from the Max Planck-uOttawa Center for Extreme and Quantum Photonics and from the International Max Planck Research School - Physics of Light (IMPRS-PL). J.G also acknowledges support from the Visiting Student Researcher Program at the uOttawa. J. G. and S.V.K acknowledge funding from the Deutsche Forschungsgemeinschaft (DFG, German Research Foundation) through Project\-ID 429529648–TRR 306 QuCoLiMa (“Quantum Cooperativity of Light and Matter”). J.B. and L.R. acknowledge funding from the Vanier Canada Graduate Scholarships program (NSERC) and the computational resources and support of the Digital Research Alliance of Canada. 


\bibliography{main}


\newpage 
\appendix
\input{sub.tex}


\end{document}

%% file: sub.tex








\title{Supplemental material}

\section{Light scattering on a magnetic sphere}
In the following we provide additional derivations for the light scattering on a magnetic sphere as discussed in the main text.

\subsection{Field expansion}
\label{sup:expansion}
All electromagnetic fields $\boldsymbol{\mathbf{F}} \in \{ \boldsymbol{\mathbf{E}}, \boldsymbol{\mathbf{B}}, \boldsymbol{\mathbf{D}}, \boldsymbol{\mathbf{H}}\}$ can be expanded as
\begin{equation}
    \boldsymbol{\mathbf{F}} (\boldsymbol{\mathbf{r}}) = \sum_{lm} \frac{1}{r} \left[ \mathcal{F}_{lm}^X(r) \boldsymbol{\mathbf{X}}_l^m (\theta, \phi) + \mathcal{F}_{lm}^Y(r) \boldsymbol{\mathbf{Y}}_l^m (\theta, \phi) + \mathcal{F}_{lm}^Z(r) \boldsymbol{\mathbf{Z}}_l^m (\theta, \phi) \right],
    \label{eq:expansion}
\end{equation}
where $\mathbf{R}_l^m$ are the vector spherical harmonics given by
\begin{equation}
    \begin{split}
        \boldsymbol{\mathbf{X}}^m_l(\theta, \phi) &= Y^m_l(\theta, \phi) \, \hat{\boldsymbol{\mathbf{r}}}, \\[0.2cm]
        \boldsymbol{\mathbf{Y}}^m_l(\theta, \phi) &= \frac{1}{\sqrt{l(l+1)}} \, \mathbf{L} \, Y^m_l(\theta, \phi) \\[0.2cm]
       &=  \frac{i}{\sqrt{l(l+1)}} \left[ \frac{1}{\sin \theta} \, \partial_{\phi} Y^m_l(\theta, \phi) \, \hat{\boldsymbol{\theta}} -  \partial_{\theta} Y^m_l(\theta, \phi) \, \hat{\boldsymbol{\phi}} \right] , \\[0.2cm]
        \boldsymbol{\mathbf{Z}}^m_l(\theta, \phi) &= \frac{ir}{\sqrt{l(l+1)}} \, \boldsymbol{\nabla} Y^m_l(\theta, \phi)  \\[0.2cm]
        & = \frac{i}{\sqrt{l(l+1)}} \left[ \partial_{\theta } Y^m_l(\theta, \phi) \, \hat{\boldsymbol{\theta}} + \frac{1}{\sin \theta} \, \partial_{\phi} Y^m_l(\theta, \phi) \, \hat{\boldsymbol{\phi}} \right]
    \end{split}
    \label{eq:vsh}
\end{equation}
with $Y^m_l$ the spherical harmonics.
Note that the factor $1/r$ is taken out of the expansion to account for far field radiation. 
Furthermore, we note that with this field expansion we are now working in spherical coordinates $\boldsymbol{\mathbf{r}} = (r, \theta, \phi)$.
Each of these components can be found using the orthonormality relation of the vector spherical harmonics
\begin{equation}
    F_{lm}^{V}(r)= r \int \sin \theta d\theta \, d\phi ~ \boldsymbol{\mathbf{F}}(\boldsymbol{\mathbf{r}})\cdot\left[\boldsymbol{\mathbf{V}}_{l}^{m}\left(\theta,\phi\right)\right]^{*}.
    \label{eq:ortho_vsh}
\end{equation}
Inserting the expansion in Eq.~\eqref{eq:expansion} in Maxwell's equations (Eq.~(3) in the main text) gives an independent set of equations for each $\{l,m\}$ pair
\begin{equation}
    \begin{split}
        D_{lm}^{Y}(r) &= \frac{1}{\omega\mu_{0}} \left[\frac{\sqrt{l(l+1)}B_{lm}^{X}(r)}{r} - i\frac{dB_{lm}^{Z}(r)}{dr} \right] \\
        B_{lm}^{X}(r) &= \frac{\sqrt{l(l+1)} E_{lm}^{Y}(r)}{\omega r} \\
        B_{lm}^{Z}(r) &= \frac{-i}{\omega}\frac{dE_{lm}^{Y}(r)}{dr} \\
        B_{lm}^{Y}(r) &= -\frac{1}{\omega} \left[\frac{\sqrt{l(l+1)} E_{lm}^{X}(r)}{r} - i\frac{dE_{lm}^{Z}(r)}{dr} \right] \\
        D_{lm}^{X}(r) &= -\frac{\sqrt{l(l+1)} B_{lm}^{Y}(r)}{\omega\mu_{0}r} \\
        D_{lm}^{Z}(r) &= \frac{i}{\omega\mu_{0}}\frac{dB_{lm}^{Y}(r)}{dr}. \\
    \end{split} \label{eq:XYZrel}
\end{equation}
Outside the sphere the permittivity is isotropic and Maxwell's equations further split into two independent sets of equations, one for ${\{E_{lm}^{Y},B_{lm}^{X},B_{lm}^{Z}\}}$ and one for ${\{B_{lm}^{Y},E_{lm}^{X},E_{lm}^{Z}\}}$. 
Outside the sphere both $E_{lm}^{Y}(r)$ and $B_{lm}^{Y}(r)$, satisfy, 
\begin{equation}
\frac{d^{2}F}{dr^{2}}+\left(k_{0}^{2}-\frac{l(l+1)}{r^{2}}\right)F=0
\label{outside_sphere}
\end{equation}
with $k_{0}=\omega/c$. 
This expression has two independent solutions known as Riccati-Bessel functions, $S_{l}(k_{0}r)$ and $C_{l}(k_{0}r)$, which are related to the Bessel functions via
\begin{equation}
    \begin{split}
        S_{l}(x) &= \sqrt{\frac{\pi x}{2}}J_{\frac{l+1}{2}}(x), \\[0.1cm]
        C_{l}(x) &= -\sqrt{\frac{\pi x}{2}}Y_{\frac{l+1}{2}}(x).
    \end{split}
\end{equation}
While $S_{l}$ is finite everywhere, $C_{l}$ diverges at the origin. The linear combinations
\begin{equation}
    \begin{split}
        \xi_{l}(x) & =S_{l}(x) - iC_{l}(x) \\[0.1cm]
        \zeta_{l}(x) &= S_{l}(x) + iC_{l}(x)
    \end{split}
\end{equation}
as $x\rightarrow\infty$ satisfy the relations
\begin{equation}
    \begin{split}
        \xi_{l}(x) &\propto e^{+ ix}, \\[0.1cm]
        \zeta_{l}(x) &\propto e^{- ix}.
    \end{split}
\end{equation}
Consequently, they can be interpreted as outgoing and incoming waves.
Given $E_{lm}^{Y}$ and $B_{lm}^{Y}$, the other components can be found from Eqs.~(\ref{eq:XYZrel}). The two sets of eigensolutions are labelled as TE (transverse electric) and TM (transverse magnetic). They are given by,
\begin{equation}
\begin{split}
    \mathbf{E}_{lm,\text{TE}} &= \frac{R_l(k_0r)}{k_0r} \mathbf{Y}_l^m(\theta,\phi) \\
    \mathbf{E}_{lm,\text{TM}} &= \frac{c}{k_0r} \left[ -\sqrt{l(l+1)} \frac{R_l(k_0r)}{k_0r} \mathbf{X}_l^m(\theta,\phi) + iR'_l(k_0r) \mathbf{Z}_l^m(\theta,\phi) \right],
\end{split} \label{eq:eigens}
\end{equation}
where $R_l(k_0r)$ can be any linear combination of $S_l$ and $C_l$. The magnetic field is given by $\text{TE}\leftrightarrow\text{TM}$, $\mathbf{E}\rightarrow\mathbf{B}$, and $c\rightarrow -1/c$.

\subsection{Incident and scattered fields}
\label{sup:incident}

The incident field should be finite everywhere as it is generated by an external source far from the sphere, so it can be modelled as the linear combination given in Eq.~(4) of the main text,
\begin{equation}
    \mathbf{E}_{\text{in}}(\mathbf{r}) = \sum_{lm} \frac{1}{k_0r} \left[ \mathcal{E}_{lm}^{\text{in}} \mathbf{E}^{\text{in}}_{lm,\text{TE}} + c\mathcal{B}_{lm}^{\text{in}} \mathbf{E}^{\text{in}}_{lm,\text{TM}}  \right], \label{eq:expansion_in}
\end{equation}
for an arbitrary set of coefficients $\{\mathcal{E},\mathcal{B}\}$ and $\mathbf{E}^{\text{in}} = k_0r\mathbf{E}$ in terms of Eqs.~(\ref{eq:eigens}) with $R_l\rightarrow S_l$.

The scattered field is given by an analogous expansion with $\mathcal{E}^{\text{in}}\rightarrow \mathcal{E}^{\text{S}}$ and $R_l\rightarrow \xi_l$ to model outgoing waves.

\subsection{Mie scattered fields}
\label{sup:mie}
When $f_{\text{F}}=0$, there is no Faraday rotation, and the problem reduces to Mie scattering. In the above notation, $\mathcal{E}_{lm}^{\text{S}} = \mathcal{E}_{lm}^{\text{M}}$.
In that case, the fields inside the sphere also satisfy Eq.~\eqref{outside_sphere} but with $k_{0}\rightarrow k=nk_{0}$ where $n$ is the refractive index. 
Thus, the fields inside the sphere ($r<R$) are given by an expansion analogous to Eq.~(\ref{eq:expansion_in}) with $\mathcal{E}^{\text{in}} \rightarrow \mathcal{E}^{\text{sp}}$ and $k_0\rightarrow k$. The coefficients can be found using the continuity of $\{D_{lm}^{X},E_{lm}^{Y},E_{lm}^{Z},B_{lm}^{X},B_{lm}^{Y},B_{lm}^{Z}\}$
for each $lm$ at the boundary. 
Explicitly,
\begin{equation}
    \begin{split}
        \mathcal{E}^{\text{sp}}_{l}S_{l}(n k_0 R) & =\mathcal{E}^{\text{in}}_{l}S_{l}(k_{0}R)+\mathcal{E}^{\text{M}}_{l}\xi_{l}(k_{0}R)\\
        k\mathcal{E}^{\text{sp}}_{l}S_{l}'(n k_0 R) & =k_{0}\mathcal{E}^{\text{in}}_{l}S_{l}'(k_{0}R)+k_{0}\mathcal{E}^{\text{M}}_{l}\xi_{l}'(k_{0}R)\\
        \mathcal{B}^{\text{sp}}_{l}S_{l}(n k_0 R) & =\mathcal{B}^{\text{in}}_{l}S_{l}(k_{0}R)+\mathcal{B}^{\text{M}}_{l}\xi_{l}(k_{0}R)\\
        \frac{k}{\epsilon}\mathcal{B}^{\text{sp}}_{l}S_{l}'(n k_0 R) & =\frac{k_{0}}{\epsilon_{0}}\mathcal{B}^{\text{in}}_{l}S_{l}'(k_{0}R)+\frac{k_{0}}{\epsilon_{0}}\mathcal{B}^{\text{M}}_{l}\xi_{l}'(k_{0}R).
\end{split}
\end{equation}
Using the identity $S_{l}(x)\xi_{l}'(x)-S_{l}'(x)\xi_{l}(x)=i$ the excitation amplitude turns out to be
\begin{equation}
    \begin{split}
        \frac{\mathcal{E}^{\text{sp}}_{l}}{\mathcal{E}^{\text{in}}_{l}} & =\frac{i}{S_{l}(n k_0 R)\xi_{l}'(k_{0}R)-nS_{l}'(n k_0 R)\xi_{l}(k_{0}R)}\triangleq\chi_{l}^{\text{E}} \label{Exc:TE}\\
        \frac{\mathcal{B}^{\text{sp}}_{l}}{\mathcal{B}^{\text{in}}_{l}} & =\frac{i}{S_{l}(n k_0 R)\xi_{l}'(k_{0}R)-n^{-1}S_{l}'(n k_0 R)\xi_{l}(k_{0}R)}\triangleq\chi_{l}^{\text{B}},
    \end{split}
\end{equation}
where $\chi$ can be interpreted as susceptibility.
The radiation amplitude of the Mie scattered field is given by
\begin{equation}
    \begin{split}
        \frac{\mathcal{E}^{\text{M}}_{l}}{\mathcal{E}^{\text{in}}_{l}} & =\frac{nS_{l}'(n k_0 R)S_{l}(k_{0}R)-S_{l}'(k_{0}R)S_{l}(n k_0 R)}{S_{l}(n k_0 R)\xi_{l}'(k_{0}R)-nS_{l}'(n k_0 R)\xi_{l}(k_{0}R)}\triangleq r_{l}^{\text{E}}\\
        \frac{\mathcal{B}^{\text{M}}_{l}}{\mathcal{B}^{\text{in}}_{l}} & =\frac{n^{-1}S_{l}'(n k_0 R)S_{l}(k_{0}R)-S_{l}'(k_{0}R)S_{l}(n k_0 R)}{S_{l}(n k_0 R)\xi_{l}'(k_{0}R)-n^{-1}S_{l}'(n k_0 R)\xi_{l}(k_{0}R)}\triangleq r_{l}^{\text{B}}.
        \label{rad_amp_mie}
    \end{split}
\end{equation}
We note that the polarization and the angular momentum are preserved in an isotropic scattering process. 

\subsection{Faraday scattered fields}
\label{sup:faraday}

When $f_{\text{F}} \ne 0$, the scattered light can be written as $\mathcal{E}_{lm}^{\text{S}} = \mathcal{E}_{lm}^{\text{M}} + \mathcal{E}_{lm}^{\text{F}}$, where $\mathcal{E}_{lm}^{\text{M}}$ was found in the previous subsection. For a sphere of radius $\SI{1}{\micro\meter}$ and typical Faraday rotation of $<1 \si{\deg/\micro\meter}$, the Faraday effect is a small perturbation. In this section, we find the leading order contribution to the scattered light due to Faraday rotation. As we show below, the Mie scattered component is largely unaffected except for a small renormalization of susceptibility $\chi$, and the Faraday effect inverts the polarization, i.e. we can write 
\begin{equation}
\begin{split} 
    \mathcal{B}_{lm}^{\text{F}} &= \frac{f_{\text{F}}}{c} \sum_L \mathcal{T}^{\text{B}}_{Llm} \mathcal{E}_{Lm}^{\text{in}} \\
    \mathcal{E}_{lm}^{\text{F}} &= f_{\text{F}}c \sum_L \mathcal{T}^{\text{E}}_{Llm} \mathcal{B}_{Lm}^{\text{in}}
\end{split} \label{def:Farcoeff}
\end{equation}
Note that the azimuthal index $m$ doesn't change due to azimuthal symmetry.

The electric field inside the sphere is given by $\mathbf{E} = \mathbf{E}_{\text{sp}} + \mathbf{E}_{\text{F}}$, where $\mathbf{E}_{\text{sp}}$ was calculated in the previous subsection and $\mathbf{E}_{\text{F}}\propto f_{\text{F}}$ is the linear correction due to the Faraday effect, to be calculated. The displacement vector is given by
\begin{equation}
    \mathbf{D} = \varepsilon_0 \left( \varepsilon_r \mathbf{E} + if_{\text{F}}\text{M}_{\text{s}} \mathbf{z}\times\mathbf{E} \right).
\end{equation}
For simplifying this further, we need the following identities
\begin{align}
    \mathbf{z}\times\mathbf{Y}_{l}^{m} &= -\frac{im}{l(l+1)}\mathbf{Y}_{l}^{m} + g_l^m\mathbf{Z}_{l-1}^{m} + g_{l+1}^m\mathbf{Z}_{l+1}^{m} - ig_l^m\sqrt{\frac{l}{l-1}}\mathbf{X}_{l-1}^{m} + ig_{l+1}^m\sqrt{\frac{l+1}{l+2}}\mathbf{X}_{l+1}^{m} \\
    \mathbf{z}\times\mathbf{X}_{l}^{m} &= -\frac{m}{\sqrt{l(l+1)}}\mathbf{Z}_{l}^{m} + ig_l^m\sqrt{\frac{l}{l+1}}\mathbf{Y}_{l-1}^{m} - ig_{l+1}^m\sqrt{\frac{l+1}{l}}\mathbf{Y}_{l+1}^{m} \\
    \mathbf{z}\times\mathbf{Z}_{l}^{m} &= -\frac{im}{l(l+1)}\mathbf{Z}_{l}^{m} - \frac{m}{\sqrt{l(l+1)}}\mathbf{X}_{l}^{m} - g_l^m\mathbf{Y}_{l-1}^{m} - g_{l+1}^m\mathbf{Y}_{l+1}^{m}.
\end{align}
Here,
\begin{equation}
    g_l^m = \sqrt{\frac{l^{2}-1}{l^{2}}\frac{l^{2}-m^{2}}{4l^{2}-1}}.
\end{equation}
These identities can be found using the known recursion relations of scalar spherical harmonics. 

\emph{TE-input:} Consider first a purely TE mode with angular momentum numbers $LM$, i.e. $\mathcal{B}_{lm}^{\text{in}} = 0$ and $\mathcal{E}_{lm}^{\text{in}} = \mathcal{E}\delta_{lL}\delta_{mM}$. For this case, we can write the VSH components of the displacement vector up to first order in $f_{\text{F}}$ as
\begin{align}
    D_{LM}^{Y} &\approx \epsilon_{s} \chi_{L}^{\text{E}}\ \text{E}_0S_{L}(kr) \left[ 1+\frac{if_{\text{F}}\text{M}_{\text{s}}}{L(L+1)} \right] + \varepsilon_0\varepsilon_r E_{F,LM}^{Y} \\
    D_{L-1,M}^{X} &\approx -\sqrt{\frac{L}{L-1}} \varepsilon_0\varepsilon_r \mathcal{I}_- S_{L}(kr) + \varepsilon_0\varepsilon_r E_{F,L-1,M}^{X} \\
    D_{L-1,M}^{Z} &\approx -i\varepsilon_0\varepsilon_r {\cal I}_{-} S_{L}(kr) + \varepsilon_0\varepsilon_r E_{F,L-1,M}^{Z} \\
    D_{L+1,M}^{X} &\approx \sqrt{\frac{L+1}{L+2}} \varepsilon_0\varepsilon_r {\cal I}_+ S_{L}(kr) + \varepsilon_0\varepsilon_r E_{F,L+1,M}^{X} \\
    D_{L+1,M}^{Z} &\approx -i\varepsilon_0\varepsilon_r {\cal I}_{+} S_{L}(kr) + \varepsilon_0\varepsilon_r E_{F,L+1,M}^{Z},
\end{align}
where ${\cal I}_{+} = g_{L+1}^M f_{\text{F}} \text{M}_{\text{s}} \chi_{L}^{\text{E}} {\cal E}$, and ${\cal I}_{-} = g_{L}^M f_{\text{F}} \text{M}_{\text{s}} \chi_{L}^{\text{E}}{\cal E}$. All the other coefficients are higher order in $f_{\text{F}}$. The corrections to $D^Y$ cause a small renormalization of the Mie frequency and can be ignored. 

We insert the above into the last three Maxwell's equations, Eqs.~\ref{eq:XYZrel}, and eliminate $E_F^X$ and $E_F^Z$ to get an equation for $B_F^Y$,
\begin{equation}
    \frac{d^{2} B_{F,L\pm1,M}^{Y}(r)}{k^{2}dr^{2}} + \left( 1 - \frac{(L\pm1+1)(L\pm1)}{k^{2}r^{2}} \right) B_{F,L\pm1,M}^{Y}(r) = \pm\frac{n {\cal I}_{\pm}}{c} S_{L\pm1}(kr),
\end{equation}
where we used the recursion relations for Riccati-Bessel functions
\begin{align}
    S_{l-1}(x) &= S_{l}'(x) + \frac{lS_{l}(x)}{x} \\
    S_{l+1}(x) &=-S_{l}'(x) + \frac{(l+1)S_{l}(x)}{x}.
\end{align}
The general solution of this differential equation is
\begin{equation}
    b_{L\pm1}^{Y}(r) = \pm\frac{n{\cal I}_{\pm}}{c} \left[p_{L\pm1}(kr)+\alpha_{\pm}S_{L\pm1}(kr)\right],
\end{equation}
where $\alpha_{\pm}$ are unknown constants and $p_{l}(x)$ is any solution satisfying
\begin{equation}
    \frac{d^{2}p_{l}}{dx^{2}}+\left(1-\frac{l(l+1)}{x^{2}}\right)p_{l}=S_{l}(x)
\end{equation}
and $p_{l}(0)=0$. While it is possible to write an explicit formula for $p_{l}$, below we need only the Wronskian $W_{l}=p_{l}'S_{l}-p_{l}S_{l}'$ which satisfies 
\begin{equation} 
W_{l}'(x)=S_{l}^{2}(x) \Rightarrow W_{l}(x) = \frac{x}{2} \left[S_{l}^{2}(x)-S_{l+1}(x)S_{l-1}(x)\right],
\end{equation}
where $p_{l}(0)=S_{l}(0)=0$ is used.

So far, we found that the electromagnetic fields inside the sphere have a large component in $\{LM,\text{TE}\}$ and small Faraday scattered components in $\{L\pm1,M,\text{TM}\}$. Via the boundary conditions, this will also hold outside the sphere, so we can write the scattered light as $\mathcal{E}_{lm} = r_L^{\text{E}} \text{E}_0\delta_{lL}\delta_{mM}$ and
\begin{equation}
    c\mathcal{B}^{\text{F}}_{lm} = f_{\text{F}}\mathcal{E}\left(P_{LM}^{\text{B}}\delta_{l(L+1)}\delta_{mM}+M_{LM}^{\text{B}}\delta_{l(L-1)}\delta_{mM}\right).
\end{equation}
We can find these coefficients by the boundary conditions that turn out to be
\begin{align}
    p_{L\pm1}(kR) + \alpha_{\pm} S_{L\pm1}(kR) &= \frac{\pm c\mathcal{B}^{\text{F}}_{L\pm1,M}}{n{\cal I}_{\pm}} \xi_{L\pm1}(k_{0}R) \\
    p'_{L\pm1}(kR) + \alpha_{\pm} S_{L\pm1}'(kR)\pm S_{L}(kR) &= \frac{\pm c\mathcal{B}^{\text{F}}_{L\pm1,M}}{{\cal I}_{\pm}} \xi_{L\pm1}'(k_{0}R).
\end{align}
These give
\begin{align}
    P^{\text{B}}_{lm} &= -i\chi_l^{\text{E}} \chi_{l+1}^{\text{B}} g_{l+1}^m \left[ W_{l+1}(kR) + S_l(kR)S_{l+1}(kR)\right] \\
    M^{\text{B}}_{lm} &= i\chi_l^{\text{E}} \chi_{l-1}^{\text{B}} g_l^m \left[ W_{l-1}(kR) - S_l(kR)S_{l-1}(kR)\right]
\end{align}

As the calculations for the TM input are analogous, we simply write down the final result here. We define similar coefficients as above,\begin{align}
    P^{\text{E}}_{lm} &= -i\chi_l^{\text{B}} \chi_{l+1}^{\text{E}} g_{l+1}^m W_{l+1}(kR) \\
    M^{\text{E}}_{lm} &= i\chi_l^{\text{B}} \chi_{l-1}^{\text{E}} g_l^m W_{l-1}(kR).
\end{align}

Then, the scattering coefficients in the notation of Eqs.~(\ref{def:Farcoeff}) are
\begin{equation}
    {\cal T}^{\sigma}_{Llm} = P_{LM}^{\sigma} \delta_{l(L+1)}\delta_{mM} + M_{LM}^{\sigma} \delta_{l(L-1)}\delta_{mM}
\end{equation}

For a plane wave, $\text{E}_0 \, e^{ik_0z} \hat{\mathbf{x}}$, we have the input~\cite{mie}
\begin{align}
    \mathcal{E}^{\text{in}}_{lm} &= \text{E}_0\sqrt{\pi(2l+1)} i^l \left(\delta_{m,1} + \delta_{m,-1}\right) \\
    \mathcal{B}^{\text{in}}_{lm} &= \frac{\mathcal{E}}{c} \sqrt{\pi(2l+1)} i^{l+1} \left(\delta_{m,1} - \delta_{m,-1}\right)
\end{align}

Then, the expressions for Mie and the Faraday scattered light, as written in the main text, directly follow.

\section{Effective permittivity tensor of magnetic materials}
For capturing the Faraday effect mathematically we derive the modified effective permittivity tensor.
We can always expand the displacement vector as
\begin{equation}
    \mathbf{D} = f_\text{E}\mathbf{E} + f_{\text{M}}\mathbf{M} + f_{\times} \mathbf{M}\times\mathbf{E},
\end{equation}
where $f$ are functions of $\mathbf{E}$ and $\mathbf{M}$. 
As the system is rotationally symmetric, all of $f$ should be a function of only $|\mathbf{E}|^2$ and $\mathbf{E}\cdot\mathbf{M}$ as $|\mathbf{M}|^2$ is a constant.
Assuming that there are no non-linear optical processes, which are weak in typical materials, we need to keep only the terms that are linear in $\mathbf{E}$. Then, the most general form is
\begin{equation}
    \mathbf{D} = \varepsilon_0 \left( \varepsilon_r \mathbf{E} + f_C \mathbf{M}\cdot\mathbf{E}\  \mathbf{M} + if_{\text{F}} \mathbf{M}\times\mathbf{E} \right),
\end{equation}
with constants $\{\varepsilon_r,f_{\text{C}},f_{\text{F}}\}$. As the permittivity should be Hermitian, all of these constants are real.

For simplicity, we ignore the Cotton-Mouton effect ($\propto f_C$). Then, we get
\begin{equation}
    \varepsilon_{ik}(\boldsymbol{\mathbf{M}}) = \varepsilon_0 \bigg( \varepsilon_r \delta_{ik} - i f_{\text{F}} \sum_j \varepsilon_{ikj} M_j \bigg).
\label{permittivity_faraday}
\end{equation} 
For including also dispersive magnetic materials the relative permittivity and the Faraday constant need to be made frequency dependent by using appropriate models for $\varepsilon_r(\omega)$ and $f_{\text{F}}(\omega)$ (see Eq.~(26) in the main text).

\section{Extension of the FDTD method to dispersive magnetic materials}
Below we provide additional derivations used for extending the FDTD method to dispersive magnetic dielectrics.

\subsection{Fourier transforms}
For transforming the optical response in the frequency domain into the time domain using a Fourier transformation, the following substitutions need to be made
\begin{equation}
    \begin{split}
        - i\omega &\rightarrow \partial_t, ~~~~~~~~~~ - \omega^2 \rightarrow \partial_t^2, \\[0.2cm]
        i \omega^3 &\rightarrow \partial_t^3, ~~~~~~~~~~~~  \omega^4 \rightarrow \partial_t^4.
    \end{split}
\end{equation}

\subsection{Update equations for the polarization vector}
For deriving the update equation for the polarization vector we need to discretize Eq.~(28) in the main text using central differencing.
By centering the central differences at time step $n-1$, the update equations for the polarization are
\begin{equation}
    \begin{split}
        \left( \frac{1}{\boldsymbol{\triangle} t^4} + \frac{\eta}{\boldsymbol{\triangle}t^3} \right) \boldsymbol{\mathbf{P}}^{n+1}
        &= \left( \frac{4}{\boldsymbol{\triangle}t^4} + \frac{2 \eta}{\boldsymbol{\triangle}t^3} - \frac{2 \omega_0^2 + \eta^2}{\boldsymbol{\triangle}t^2} - \frac{\omega_0^2 \eta}{\boldsymbol{\triangle}t}  \right) \boldsymbol{\mathbf{P}}^n + \left( -\omega_0^4 - \frac{6}{\boldsymbol{\triangle}t^4} + 2 \frac{2 \omega_0^2 + \eta^2}{\boldsymbol{\triangle}t^2} \right) \boldsymbol{\mathbf{P}}^{n-1} \\[0.2cm]
        &+ \left( \frac{4}{\boldsymbol{\triangle}t^4} - \frac{2 \eta}{\boldsymbol{\triangle}t^3} - \frac{2 \omega_0^2 + \eta}{\boldsymbol{\triangle}t^2} \frac{\omega_0^2 \eta}{\boldsymbol{\triangle}t} \right) \boldsymbol{\mathbf{P}}^{n-2} + \left( - \frac{1}{\boldsymbol{\triangle}t^4} + \frac{\eta}{\boldsymbol{\triangle}t^3} \right) \boldsymbol{\mathbf{P}}^{n-3} \\[0.2cm]
        & + \omega_0^2 \varepsilon_0 (\varepsilon -1 ) \Big[ \left( \frac{1}{\boldsymbol{\triangle}t^2} + \frac{\eta}{2\boldsymbol{\triangle}t} \right) \boldsymbol{\mathbf{E}}^n  + \left( \omega_0^2 - \frac{2}{\boldsymbol{\triangle}t^2} \right) \boldsymbol{\mathbf{E}}_{n-1}\\[0.2cm]
        &  + \left( \frac{1}{\boldsymbol{\triangle}t^2} - \frac{\eta}{2 \boldsymbol{\triangle}t} \right) \boldsymbol{\mathbf{E}}^{n-1} \Big] - \frac{\varepsilon_0 A_3 \omega_0}{2 \boldsymbol{\triangle} t} \, \boldsymbol{\boldsymbol{\mathbf{M}}} \times \left( \boldsymbol{\mathbf{E}}^n - \boldsymbol{\mathbf{E}}^{n-2} \right)
    \end{split}
\end{equation}
which then include the magnetic model into the update equations via the electric field
\begin{equation}
    \boldsymbol{\mathbf{E}}^{n+1} = \boldsymbol{\mathbf{E}}^{n} + \frac{\boldsymbol{\triangle} t}{\varepsilon_0} \, \nabla \times \boldsymbol{\mathbf{H}}^{n+1/2} - \frac{1}{\varepsilon_0} \left( \boldsymbol{\mathbf{P}}^{n+1} - \boldsymbol{\mathbf{P}}^n \right).
\end{equation}
Thus, for any given time step, we need to store $\boldsymbol{\mathbf{E}}^{n+1}$, $\boldsymbol{\mathbf{E}}^n$, $\boldsymbol{\mathbf{E}}^{n-1}$, $\boldsymbol{\mathbf{E}}^{n-2}$, $\boldsymbol{\mathbf{P}}^{n+1}$, $\boldsymbol{\mathbf{P}}^n$, $\boldsymbol{\mathbf{P}}^{n-1}$, $\boldsymbol{\mathbf{P}}^{n-2}$, and $\boldsymbol{\mathbf{P}}^{n-3}$.

